# DQE as quantum efficiency of imaging detectors

*Giovanni Zanella*

*Dipartimento di Fisica "Galileo Galilei", Università di Padova and
I.N.F.N.-Sezione di Padova,
via Marzolo, 8  -I-35131 Padova, Italy*

**Abstract**

In this paper a general and unified treatment of the DQE is exposed, both in the space-domain and in the frequency-domain. The meaning of the DQE as quantum efficiency appears to be of paramount importance for a correct interpretation of the physical parameters involved in its formulation and measurement. The treatment in the frequency-domain turns out to be a direct extension of that in the space-domain.
An operational procedure is introduced to consider the effect of the filtering of the detector in the space-domain.

## 1. Introduction

The detective quantum efficiency (DQE) is a parameter introduced to assess the varying levels of performance of imaging detectors with the aim of comparing their imaging capabilities by an unified approach [1][2].

Despite its being widespread, this parameter is not generally well understood and consequently it is not well measured or used.

As we shall see, various and subtle problems appear in the concept, the formulation, and in the appropriate use of the DQE.

The words "quantum efficiency" have a precise meaning, because the DQE measures the quantum efficiency of an *equivalent virtual detector* which produces at its output the same signal-to-noise ratio of the real detector even when the input signal is the same.

In this context the *quantum efficiency* (QE) of a detector is intended as the *average fraction of the input quanta which is used in the formation of the output signal*, no matter if the single input quantum generates a distinct output signal or not.

The DQE is generally defined by the ratio of the *squared output signal-to-noise ratio* $(SNR_o)^2$ to the *squared input signal-to-noise ratio* $(SNR_i)^2$ of the imaging detector. This definition of the DQE, in accordance with the exact meaning of the terms used, is, however, a consequence of its primary meaning of quantum efficiency.

The discussion which follows will be carried out at first within the *space-domain*, since the DQE treatment in the spatial *frequency-domain* is an extension of concepts established in the more familiar *space-domain*.

The DQE formulation in the spatial *frequency-domain* is useful for highlighting the detector capabilities at a done spatial frequency for a done input signal. The DQE expressed in the *space-domain* reveals the efficiency of the detector as depending on the input signal. In any case, as we shall see, the DQE expressed in the space-domain can also be affected by the spatial resolution of the detector.

## 2. The quasi-ideal imaging detector

A good understanding of the DQE is impossible without having as a reference the "virtual" detector termed *quasi-ideal* by R.C. Jones [2]. This *quasi-ideal* imaging detector only has noise due to the input Poisson statistics and to the fluctuations introduced by the binomial statistics due to a QE < 1 [3]. These two noises are not correlated, so the noise variance at the output of a *quasi-ideal* imaging detector will be:

$$s_o^2 = (s_i)_o^2 + s_{QE}^2 \quad , \tag{1}$$

$(s_i)_o^2$ represents the input noise variance $s_i^2$ as viewed at the detector output, and $s_{QE}^2$ is the variance introduced by the binomial fluctuations due to *QE < 1*. Then

$$(s_i)_o^2 = QE^2 s_i^2 = QE^2 \overline{S_i} \quad \text{and} \quad s_{QE}^2 = QE(1-QE)\overline{S_i} \quad , \tag{2}$$

$\overline{S_i}$ being the *average input signal* expressed in quanta, which is the exact variance due to the input Poisson fluctuations.
Thus

$$s_o^2 = QE^2 \overline{S_i} + QE(1-QE)\overline{S_i} = QE\overline{S_i} = \overline{S_o} \quad , \tag{3}$$

where $QE\overline{S_i}$ denotes the average output signal $\overline{S_o}$, which is also the variance due to a Poisson statistics.
If QE = 1, $s_o^2 = s_i^2 = \overline{S_i} = \overline{S_o}$.
Supposing the ergodicity of the various signals and noises, any single measurement can be executed irrespective of the space domain or the time domain. In other words, if the signals refer to a single pixel, the averaging operation and the fluctuations from the mean must be conceived within the time. On the contrary, when we operate only on a single frame the averaging operation and the variance measurement are intended among the various pixels.
The introduction of the squared SNRs allows us to discover a *key property* of the *quasi-ideal* imaging detectors. In fact

$$\left(\frac{\overline{S_o}}{s_o}\right)^2 = \frac{QE^2 \overline{S_i}^2}{QE \overline{S_i}} = QE \overline{S_i} = QE \frac{\overline{S_i}^2}{\overline{S_i}} = QE \left(\frac{\overline{S_i}}{s_i}\right)^2 , \qquad (4)$$

so

$$QE = \frac{\left(\frac{\overline{S_o}}{s_o}\right)^2}{\left(\frac{\overline{S_i}}{s_i}\right)^2} = \frac{\overline{S_o}}{\overline{S_i}} . \qquad (5)$$

These elementary operations are very important because they contain the concept of DQE.

It is necessary to keep in mind that in the equation (5) the noises are intended as fluctuations (in quanta) with respect to an average summation of input quanta ($\overline{S_i}$) or of output quanta $\overline{S_o}$, provided $\overline{S_o} = QE \, \overline{S_i}$.

## 3. The real imaging detector

A real imaging detector has adjunctive noise sources (of variance $s_n^2$) and adjunctive backgrounds, compared to a *quasi-ideal* detector. In this case, being as the noises are not correlated,

$$s_o^2 = (s_i)_o^2 + s_{QE}^2 + s_n^2 . \qquad (6)$$

The QE of a real detector is the same as the corresponding *quasi-ideal* detector. In fact, to obtain $\overline{S_o}$, the backgrounds must be subtracted from the average output because they are steady quantities. Instead, the noises do not interfere with $\overline{S_o}$ because they have zero means.
Therefore, for a real detector

$$QE = \frac{\overline{S_o}}{\overline{S_i}} , \qquad (7)$$

and thanks to equation (3),

$$s_o^2 = \overline{S_o} + s_n^2 , \qquad (8)$$

where in this case $\overline{S_o}$ is a variance measured by quanta$^2$.
Consequently, our real detector has a smaller $SNR_o$ than that of the corresponding *quasi-ideal* detector (which has same QE and same $SNR_i$).

We can conclude that the equation (5) cannot represent the QE of a real detector, but rather the QE of the equivalent *quasi-ideal* detector which has same $SNR_o$ and same $SNR_i$ as the real detector. This *equivalent* QE is the DQE:

$$DQE = \frac{\left(\dfrac{\overline{S_o}}{s_o}\right)^2}{\left(\dfrac{\overline{S_i}}{s_i}\right)^2} \quad . \qquad (9)$$

In practice, the performance of any real imaging detector is translated by DQE into that of the equivalent *quasi-ideal* detector, so the imaging capabilities of different detectors can be effectively compared. Obviously, QE and DQE of *quasi-ideal* detectors are coincidental.

Using the equation (8), DQE can be written in the following manner:

$$DQE = \frac{\dfrac{\overline{S_o}^2}{\overline{S_o} + s_n^2}}{\overline{S_i}} = \frac{QE}{1 + \dfrac{s_n^2}{QE\,\overline{S_i}}} \quad . \qquad (10)$$

Unlike QE, the DQE is not a simple constant characteristic of the detector, and depends not only on QE, but also on the average input signal $\overline{S_i}$ (the variance of the input Poisson noise) and on the variance $s_n^2$ of the adjunctive noise of the imaging detector.

From equation (10) stems the relation:

$$DQE \leq QE \leq 1 \quad . \qquad (11)$$

It can be demonstrated that the knowledge of the DQE is equivalent to the knowledge of other parameters of an imaging detector like the *detectable minimum signal difference*, the *maximum number of resolvable grey levels*, and the *detectable minimum contrast* [4].

## 4. Numerical examples

To be able to see the behaviour of imaging detectors in various experimental cases, Fig.1 displays DQE graphs obtained using equation (10) and involving four different possibilities, assuming that the saturation value of $\overline{S_i}$ is $10^8$ quanta.

Curve A considers the case of QE = 0.8 and $s_n^2 = 10^5$ quanta$^2$. Curve B pertains a detector with the same QE, but with $s_n^2 = 10$ quanta$^2$. Curve C treats the case of QE = 0,3 with the value of $s_n^2 = 0.1$ quanta$^2$ and curve D the case of QE = 1 with $s_n^2 = 10^3$ quanta$^2$.

These plots indicate that at high input signals the DQE is determined only by QE while, at low $\overline{S_i}$, the adjunctive detector noise plays a relevant role.

Plot C of Fig.1 is emblematic insofar as at low $\overline{S}_i$ the detector has the best performance, while at high $\overline{S}_i$ the DQE reaches the steady value of QE = 0.3 (the worst performance).

Concluding, in the interest of obtaining the highest DQE, we can operate on three parameters: QE, $s_n^2$ and $\overline{S}_i$, as shown in equation (10). In any case, the improvement of QE is always convenient if we maintain $s_n^2$ steady (see plot B vs. plot D).

Conversely, at low values of $\overline{S}_i$ the lowering of $s_n^2$ can dominate (improving the DQE) in the presence of a contemporary lowering of QE (see plot C vs. plots A, B, D).

In practice, the selection of the right detector requires the preliminary knowledge of $\overline{S}_i$, that is the rate of the input quanta and the integration time. Thereafter, once $\overline{S}_i$ is known we can verify if the DQE of our detector is satisfactory. On the contrary, it is interesting to see in Fig.1 how the comparison of different imaging systems performed only at a single value of $\overline{S}_i$ can be misleading.

## 5. The single quantum detection

It is a common conviction that a detector with sensitivity heightened to detect a single input quantum has the best DQE, especially if QE = 1. This opinion ignores the fact that an image is built by a summing of events, or that the effect of the "additional" noises which determine $s_n^2$ can influence the DQE, as shown by equation (10).

The detection of the single input quanta is necessary only when a decision must be reached on the basis of the intensity or timing of the signal generated by the single event. In any case, the detection of single quanta in "counting" detectors can limit the rate of the detector itself, because generally in this type of detectors the concomitant events cannot be resolved spatially.

When a "counting" detector detects a fraction of the input quanta during an integration time, $s_n^2$ derives from the fluctuations of the "spurious" events produced by spikes of the "additional" noise which exceed the acceptance threshold.

When the threshold level is maintained high, to avoid the "spurious" events, a QE loss may appear due to the intrinsic fluctuations of the single event signals. In this latter case $s_n^2 = 0$, and the DQE = QE for any $\overline{S}_i$.

The variance $s_n^2$ can be measured in the "counting" detectors by determining the variance of the fluctuations of the output signal when the input quanta are missing.

The counting of the single events is also possible using "integrating" detectors if the spots generated at the detector output by the single input quanta can be detected and spatially resolved within a frame by a pattern recognition procedure. A further image analysis is necessary to determine the pixel which is the "centroide" of these spots. The final image is obtained using these pixels as counts by adding together the content of consecutive frames.

## 6. How does an image intensifier work?

Contrary to its name, an image intensifier does not always improve the DQE of an imaging detector. In some circumstances, the use of an image intensifier can worsen the DQE of the detector, especially at high values of $\overline{S}_i$.

For example, the C curve in Fig.1 can pertain the DQE of an intensified imaging detector, where the image intensifier is used with the aim of raising the DQE at low $\overline{S}_i$. In this detector a QE of 0.3 is likely because the image intensifier has a photocathode at its input and, as we know, the quantum efficiency of the photocathode is low.

Numerical examples of DQE calculations in intensified, or non intensified, CCD imaging detector can be found in reference [6].

With reference to equation (10), the raising of the DQE in spite of the lowering of QE and of $\overline{S}_i$ requires the fall of $s_n^2$. The image intensifier operates in exactly the same way reducing $s_n^2$ (expressed in quanta$^2$) this is because it "intensifies" the signal due to the single input quantum.

## 7. The analog integration of events

When signals and noises are integrated in analog form in an "integrating" detector, the variance of the "adjunctive" noise is

$$s_n^2 = QE\, \overline{S}_i\, s_{intr}^2 + s_d^2 \quad , \qquad (12)$$

where $s_{intr}^2$ represents the variance due to the "intrinsic" fluctuations of the output signal generated by the single input quantum, and where $s_d^2$ is the variance of other "adjunctive" detector noises.

Using equation (12), equation (10) becomes:

$$DQE == \frac{QE}{1 + s_{intr}^2 + \dfrac{s_d^2}{QE\, \overline{S}_i}} \quad . \qquad (13)$$

Therefore, in the "integrating" detectors we always have DQE < QE.

The variance of the fluctuations of the output signal, without the presence of input quanta, allows us to measure $s_d^2$.

The measurement of $s_{intr}^2$ is possible if *the detector can detect the single input quantum* determining the variance of the fluctuations of the intensity of the signals due to the single detected quanta and subtracting $s_d^2$ [5].

If *the detector cannot detect the single input quantum*, we can measure $s^2_{intr}$ by determining $s^2_o$ in the presence of the input quanta and $s^2_d$ without the input quanta, then by using the equation (8):

$$s^2_{intr} = \frac{s^2_n - s^2_d}{QE\, \overline{S_i}} = \frac{s^2_o - QE\, \overline{S_i} - s^2_d}{QE\, \overline{S_i}} \quad . \tag{14}$$

If we suppose a pure Poisson statistics $s^2_{intr} = 1$ quanta$^2$ and DQE $\leq 0.5$ QE.
In conclusion, the analog integration of the events is not a convenient procedure because the DQE suffers from the presence of $s^2_{intr}$ and $s^2_d$.
The best imaging procedure using an "integrating" detector would be the analysis of the spots (if detected) generated by the input quanta, as described in paragraph 5 and practically using the "integrating" detectors as "counting" detectors.

## 8. Spatial resolution and DQE of the detector

The equation $\overline{S_o} = QE\, \overline{S_i}$ is a "quantum-equation" because $\overline{S_o}$ represents how much of the input quanta $\overline{S_i}$ are detected, apart from the average area (pixel$^2$) covered by the signals relating to $\overline{S_o}$. This area is the circular portion of the detector output frame determined by scattering and/or diffusion phenomena due to the secondary products generated by each single input quantum.
$\overline{S_i}$ is generally referred to one pixel, so the average area $\overline{n}$ (pixel$^2$) covered by the events pertaining $\overline{S_o}$ represents the minimum area which can be resolved by the detector. This area is very important because it determines the value of $s^2_n$ and of $s^2_d$.
In fact, Fig. 2 clearly shows, if $(s^2_n)_{pix}$ denotes the variance of the "adjunctive" noise per pixel in the "counting" detectors and $(s^2_d)_{pix}$ the variance of the "adjunctive" noise per pixel in the "integrating" detectors

$$s^2_n = \overline{n}\,(s^2_n)_{pix} \quad \text{and} \quad s^2_d = \overline{n}\,(s^2_d)_{pix} \quad . \tag{15}$$

Therefore, SNR$_o$, and then DQE depend on $\overline{n}$.
The average area $\overline{n}$ is related to the spatial resolution of the detector expressed by the MTF, which is the module of the *Optical Transfer Function* (OTF) of the system [7].
Fig.2 shows the line pairs of a square-wave pattern placed at the input of the detector and as it is viewed at the output when the line pairs are barely resolved sufficiently.
The line pairs of Fig.2 have the frequency limit $f_{lim}$, that is the maximum number of line pairs per millimetre (lp/mm) which can be spatially resolved by the detector. In fact, the circular spot of average area $\overline{n}$ contains exactly one pair of these lines of frequency $f_{lim}$.

The square-wave pattern pertains the so-called Contrast Transfer Function (CTF) instead of the sine-wave pattern which connects the MTF, but for our purposes the result obtained by the CTF is practically the same as that by the MTF [8].

The spatial frequency $f_{lim}$ corresponds to that in which the CTF (or the MTF) reaches the zero value. Therefore, if pixel of the detector are squares (the pixel size $Dx = Dy$), the average area $\bar{n}$, will be:

$$\bar{n} = \frac{p}{4 f_{lim}^2 \Delta x^2} \quad (pixels^2) \quad , \tag{16}$$

where $Dx$ is measured in millimetres.

## 9. The DQE in the frequency-domain

The passage from the *space-domain* to the *frequency-domain* can take place by the introduction of the *power spectral density*, or simply the *power spectrum* (also called the *Wiener spectrum*), which is the average power of the signal (or of the noise) in a unitary bandwidth centred at the frequency $f$.

Therefore, we can decide the calculation or the measurement of the DQE in function of the spatial frequencies $f_x, f_y$ along the $x, y$ axis using the power spectra.

It is convenient to write equation (9) according to equation (8) and (10) in the following manner:

$$DQE(f) = \frac{\dfrac{\overline{S_o}^2}{\overline{S_o + s_n^2}}}{\dfrac{\overline{S_i}^2}{\overline{S_i}}} = \frac{\dfrac{QE^2 \overline{S_i}^2}{QE \overline{S_i} + s_n^2}}{\dfrac{\overline{S_i}^2}{\overline{S_i}}} \quad . \tag{17}$$

Now, indicating the two variables $f_x, f_y$ simply by $f$ (often the variable $f$ represents only the frequency $f_x$) and delimiting the bandwidth to an unitary interval centred on the frequency $f$, equation (17) becomes:

$$DQE(f) = \frac{\dfrac{QE^2 W_i(f) MTF(f)^2}{QE\, W_{ni}(f) MTF(f)^2 + W_n(f)}}{\dfrac{W_i(f)}{W_{ni}(f)}} \quad , \tag{18}$$

where $W_i(f)$ and $W_{ni}(f)$ are the signal and the noise power spectra at the input of the detector and $W_n(f)$ the power spectrum of the "adjunctive" noises.

The $MTF^2$ of equation (18) permits the filtering of the power spectra $W_i(f)$ and $W_{ni}(f)$ through the system.

Equation (18) can become:

$$DQE(f) = \frac{QE}{1 + \dfrac{W_n(f)}{QE\, W_{ni}(f)\, MTF(f)^2}} \quad . \tag{19}$$

In the case of a flat field input image with Poisson fluctuations and assuming single-sided spectra [9]:

$$W_{ni}(f_x, f_y) = 4\overline{S_i}\, \frac{\sin^2(pf_x \Delta x)}{(pf_x \Delta x)^2}\, \frac{\sin^2(pf_y \Delta y)}{(pf_y \Delta y)^2} \quad , \tag{20}$$

where the *sinc²* terms are filtering terms due to the sampling array ($W_n$ and $\overline{S_i}$ are ascribed to the pixel). So we can introduce $MTF(f)|_{tot}$:

$$MTF(f_x, f_y)|_{tot} = \frac{\sin(pf_x \Delta x)}{(pf_x \Delta x)}\, \frac{\sin(pf_y \Delta y)}{(pf_y \Delta y)^2}\, MTF(f_x, f_y) \quad , \tag{21}$$

where $\Delta x$, $\Delta y$ denote the pixel sizes and MTF(f) is the so-called "presampling MTF". Therefore equation (20) becomes:

$$DQE(f) = \frac{QE}{1 + \dfrac{W_n(f)}{4QE\, \overline{S_i}\, MTF(f)|_{tot}^2}} \tag{22}$$

This equation is the same as the one in reference [9] and the DQE(0) is the so-called "zero spatial-frequency DQE" being $MTF(0)|_{tot} = 1$.

## 10. Conclusions

In this paper the DQE of the imaging detectors has been treated under the point of view of a quantum efficiency, both in the space-domain and in the frequency-domain. The meaning of quantum efficiency of the DQE is essential for a correct interpretation of the involved parameters in its formulation and measurement.
The treatment of the DQE in the frequency-domain turns out to be a consequence of that in the space-domain.
An operational procedure has been exposed to consider the filtering of the detector also in the space-domain.

**Figure captions**

Fig.1. Examples of different DQE plots (see text).

Fig.2. Single pixel spread in an imaging detector, due to scattering and /or diffusion phenomena, vs. the maximum number of spatially resolved line pairs per millimetre using a square-wave bar pattern placed at the detector input (see text).

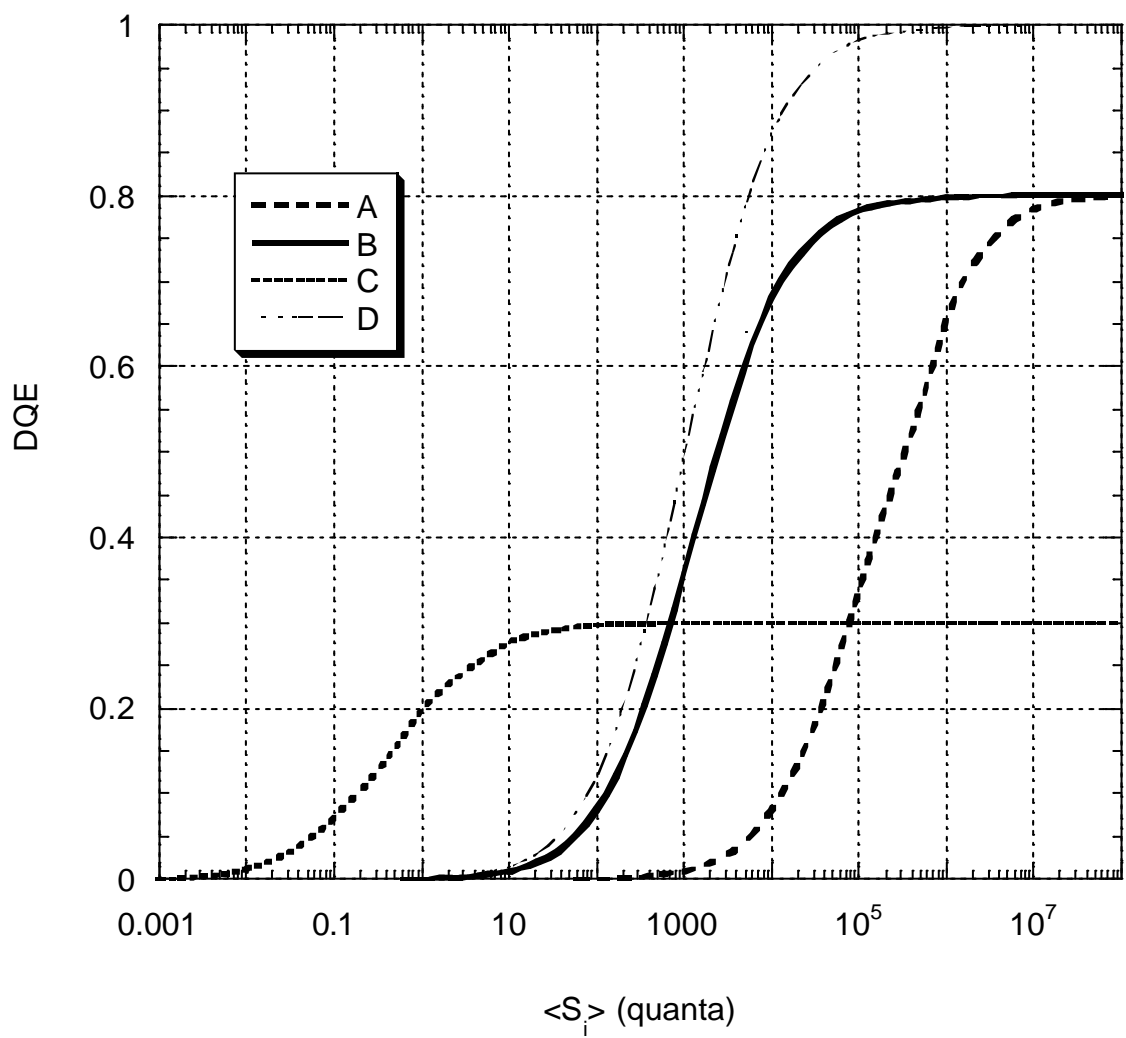

**Fig.1**

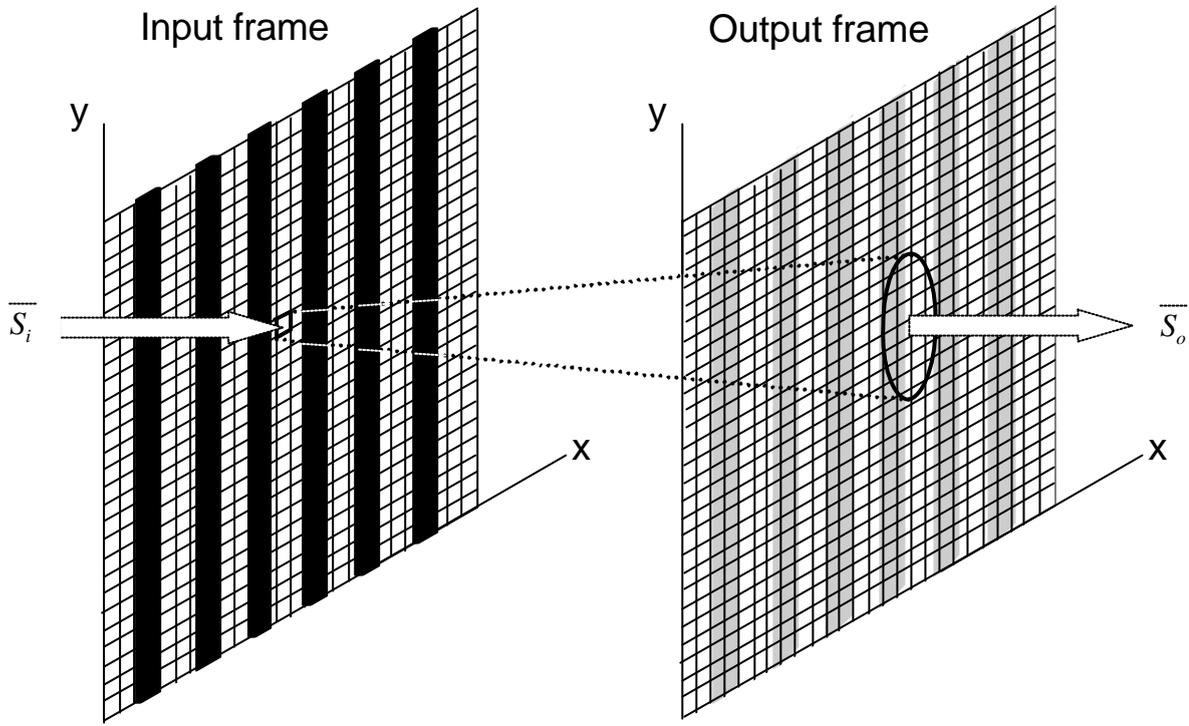

**Fig.2**